# Rethinking the Role of Pre-ranking in Large-scale E-Commerce Searching System

Zhixuan Zhang
Yuheng Huang*
Dan Ou
zhibing.zzx@alibaba-inc.com
hexing.hyh@alibaba-inc.com
oudan.od@alibaba-inc.com
Alibaba Group
Hangzhou, China

Sen Li
Longbin Li
Qingwen Liu
Xiaoyi Zeng
lisen.lisen@alibaba-inc.com
lilongbin.llb@alibaba-inc.com
xiangsheng.lqw@alibaba-inc.com
yuanhan@taobao.com
Alibaba Group
Hangzhou, China

## ABSTRACT

Large-scale e-commerce search systems such as Taobao Search, the largest e-commerce searching system in China, aim at providing users with the most preferred items (e.g., products). Due to the massive amount of data and the need for real-time responses, a typical industrial ranking system often consists of three or more modules, including matching, pre-ranking, and ranking. The pre-ranking is widely considered a mini-ranking module, as it needs to rank hundreds of times more items than the ranking under the same time latency. Existing researches focus on building a lighter model that imitates the ranking model's capability. As such, the metric of a pre-ranking model usually follows the ranking model using Area Under ROC (AUC) for offline evaluation. However, such a metric is inconsistent with online A/B tests in practice, so researchers have to perform costly online tests to reach a convincing conclusion. In our work, we rethink the role of the pre-ranking. We argue that the primary goal of the pre-ranking stage is to return an optimal unordered set rather than an ordered list of items because the ranking expert on it mainly determines the final exposures. Since AUC measures the quality of an ordered item list, it is not suitable for evaluating the quality of the output unordered set. This paper proposes a new evaluation metric called All-Scenario Hitrate (ASH) for pre-ranking. ASH is proven effective in the offline evaluation and consistent with online A/B tests based on numerous experiments in Taobao Search. To reach our proposed goals, we introduce an all-scenario-based multi-objective learning framework (ASMOL), which improves the ASH significantly. Surprisingly, the new pre-ranking model outperforms the ranking model when outputting thousands of items, which is contradictory to common sense. The phenomenon validates that the pre-ranking stage should focus on outputting thousands of items with higher quality instead of imitating the ranking blindly. With the improvements in ASH consistently translating to online improvement, it further validates that the ASH is a more effective offline metric and makes a 1.2% GMV improvement on Taobao Search.

## 1 INTRODUCTION

Taobao Search, the largest e-commerce searching system in China, is a typical multi-stage e-commerce ranking system, mainly consisting of matching, pre-ranking, ranking, and re-ranking, as is illustrated in Figure 1. The matching (also called retrieval) stage typically consists of diverse types of modules, such as textual matching, knowledge graph-based matching, and personalized embedding-based matching. It aims to output hundreds of thousands of items (i.e., products) of high quality (i.e., user-preferred items) from billions of ones. The pre-ranking stage needs to select thousands of items from the matching output set. Then it will go through a ranking and a re-ranking for the final selection and expose the best tens of items for users.

The pre-ranking is widely considered a mini-ranking module with lighter network architecture and fewer features. It needs to rank hundreds of times more items than the ranking under the same time latency. Under the strict latency constraint, the Deep Neural Network (DNN) used in pre-ranking is usually a simple vector-product-based model [8]. It consists of one user tower and one item tower, which are calculated individually. The final score is the cosine distance between the user and item vectors. Therefore, the ability of the vector-product-based model is relatively weaker than that of the downstream ranking model that usually firstly concatenates user and item features and feeds them into a wide and deep fully-connected network [4].

*Corresponding author

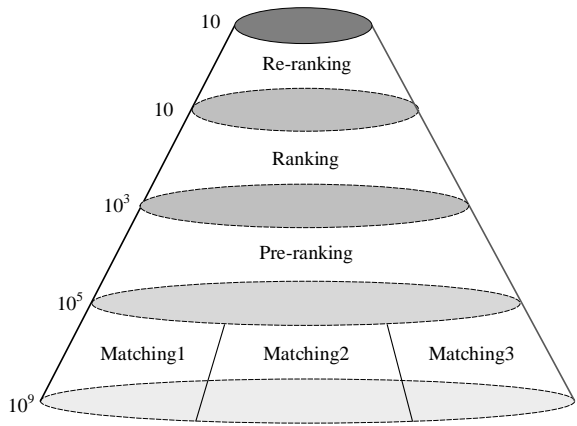

**Figure 1: The multi-stage ranking system in Taobao Search**

In recent years, most researchers have focused on building a lighter model that imitates the ranking via feature selection, network compression, knowledge distillation, etc. [11, 15, 24]. After a long period of practice, we find that imitating the ranking can improve the consistency between the ranking and the pre-ranking. As such, the high-quality items selected by the pre-ranking are more likely to be exposed by the ranking. Although these methods can improve online business metrics as a short-term solution, they rarely benefit the overall item quality of the pre-ranking stage. In Taobao Search, these optimizations can only make the pre-ranking model output a few more high-quality items among the thousands of output items. Nevertheless, it can hardly benefit the entire multi-stage system in the long-term view. Because the pre-ranking in Taobao Search selects items from hundreds of thousands of candidates, only using exposures (also called impressions [7]) during training as the ranking does, will lead to a severe Sample Selection Bias (SSB) problem [25]. Specifically, the item features distribution in exposures and the pre-ranking candidates are quite different. As a result, most items' scores in the pre-ranking candidates can be unconvincing because the model rarely learns it during training. Furthermore, although these methods can be effective in a short period, or when the number of candidates in pre-ranking does not far exceed that in the ranking, it indeed results in the Matthew Effect and is detrimental to the entire search system in a long-term perspective.

Besides the SSB problem in the pre-ranking stage, we also suppose that the goals of the ranking and the pre-ranking are different. The ranking, which is expert on selecting the top items, will re-rank the order inside the pre-ranking outputs and determine the final output. As a result, **the primary goal of the pre-ranking stage should be to return an optimal unordered set rather than an ordered list of items.** Based on the analysis above, together with online and offline experiments on Taobao Search, we rethink the role of a pre-ranking and re-define two goals for pre-ranking:

- **High quality set**: to improve the quality of the output set by solving the SSB on pre-ranking candidates.

- **High quality rank**: to get consistent with ranking on the output set, ensuring that high-quality items can get high scores in ranking and be exposed to users.

However, **it is impossible to maximize both two goals simultaneously**. The first goal plays a dominant role between the two goals and is the goal a pre-ranking model must pursue. As for the second goal, we only need to satisfy it when ensuring the quality of the output set does not drop. In other words, when the model reaches the Pareto frontier, there is usually a "Seesaw Effect" between the quality of the whole output set and its inside rank. The rank inside its output set will get worse when it focuses more on the whole set without involving more online computation. Similarly, the quality of the whole output set drops when imitating the ranking and improving its inside AUC without involving more online computation. It is why AUC is not consistent with online business metrics. We will detail this phenomenon in Section 4.

The existing offline evaluation metrics like Area Under ROC (AUC) [24] can measure pre-ranking ability corresponding to the second goal. However, AUC measuring the quality of an ordered item list is not suitable for evaluating the quality of the output unordered set. And there is no metric that can effectively measure the first goal. Although there are a few researchers in the industry that try to improve the quality of the output set, they have not provided a reliable offline evaluation metric to measure the improvement. In practice, the most common strategy is measuring the improvement by online A/B testing. However, the online evaluation is costly and time-consuming because it usually takes several weeks to get a convincing result. In this paper, we propose a new evaluation metric, called All-Scenario Hitrate (ASH), for measuring the quality of a pre-ranking model's outputs. Through a systematic analysis of the relationship between ASH and the online business metric, we validate the effectiveness of this new offline evaluation metric. To reach our proposed goals, we further propose an all-scenario-based multi-objective learning framework (ASMOL) which improves the ASH significantly. Surprisingly, the new pre-ranking model can outperform the ranking model when outputting thousands of items. **The phenomenon further validates that the pre-ranking stage should focus on outputting higher quality set and should not imitate the ranking blindly.** With the improvements in ASH consistently translating to online improvement, it further validates that the ASH is a more effective offline metric and makes a 1.2% GMV improvement on Taobao Search.

To sum up, our main contributions are threefold:

- rethinking and re-defining the two goals of pre-ranking, proposing the corresponding evaluation metric, and demonstrating how and why the new metric ASH works by analyzing each stage on Taobao Search.
- proposing an all-scenario-based multi-objective learning framework (ASMOL) to improve the two offline evaluation metrics, which aims to reach the proposed two goals of pre-ranking.
- showing the correspondence between ASH and the online business metrics on Taobao Search via massive online experiments and eventually improving 1.2% GMV on Taobao Search.

## 2 RELATED WORK

The number of candidate items in the industrial e-commerce searching system is enormous. For example, there are billions of candidate items in Taobao Search. With so many candidate items, it is common to use a cascade ranking system [5, 6, 14] to pick out the most preferred items. Generally, the cascade ranking system typically consists of three stages: matching, pre-ranking, and ranking.

### 2.1 Pre-ranking

In a large-scale e-commerce multi-stage cascade ranking system, the pre-ranking has long played the role of a mini-ranking module. A pre-ranking receives hundreds of thousands of candidates from the matching stage and feeds the best top thousands to the ranking stage. Since the number of candidate sets in the pre-ranking is tens or even hundreds of times larger than that in the ranking stage, the pre-ranking has to make a trade-off between effectiveness and efficiency. Therefore, most research focuses on improving the ranking ability of the pre-ranking model as much as possible while ensuring low latency.

The vector-product-based DNN model [8] is widely-favored in industrial pre-ranking modules. The vector-product-based model map query and item to two vectors with two separate DNNs, and compute the cosine similarity between the query vector and the item vector as the ranking score. Since the item features are usually stable, the item vector can be pre-computed and cached, which significantly reduces the overhead in online serving. On this basis, FSCD [15] focused on using more effective features with a learnable feature selection approach. RD [23] learns a compact ranking model with knowledge distillation. To further improve the model ability, COLD [24] use a fully-connected layer to replace the cosine similarity, and reduce the computational cost by feature selection and engineered optimization. To find a more reasonable network structure and features, AutoFAS [11] introduces network architecture searching (NAS) to find the optimal network structure and features automatically. The main contributions of all these works focus on improving the ranking ability of the model while ensuring low latency, and the most popular evaluation metrics are AUC [24] and $hitrate@k$ ($recall@k$) [11].

### 2.2 Ranking and Matching

The ranking module is designed to rank the thousands of candidate items fed by the pre-ranking module, and output dozens of items to the user interface. Most researchers of ranking focus on network structure and the user behaviour modeling, such as modeling user behavior with the attention mechanism [3, 16, 30], feature interaction [12, 21] and representative learning [2, 13]. However, recent works in these areas are difficult to be directly learned by the pre-ranking stage because they introduce increasing computational complexity, which is unacceptable in the pre-ranking stage. Therefore, a great deal of attention is paid to model efficiency in the previous pre-ranking researches[11, 15].

The matching stage retrieves hundreds of thousands of candidates from billions of candidates and feeds into the pre-ranking stage. In the industrial search system, there are always multiple matching algorithms. The results of all matching algorithms are merged and fed into the pre-ranking module. Common matching methods include semantic matching [9], behavior-based matching [18], embedding-based matching [7, 10], multi-modal matching [28], etc. The embedding-based matching trains embedding vectors for both users and items, and retrieves the item vectors with the approximate nearest neighbor algorithm at online serving. The embedding training in the embedding-based matching and the vector-product-based pre-ranking is similar. While most embedding-based matching research points out the importance of the sample selection [6, 7, 26], we notice that previous pre-rank researches paid to the sample selection strategy. Almost all existing pre-ranking models blindly follow exactly the same training samples as the ranking model. In Section 4, we will thoroughly discuss the sample selection strategies in the pre-ranking stage.

The three stages in the cascade ranking system are closely related. From the view of the sample, the candidates of the pre-ranking stage depend on the matching stage. From the view of the model structure, the pre-ranking models widely inherit the feature engineering and user behavior modeling techniques in the ranking stage, and the vector-product-based structure is widely used in both the matching and the pre-ranking. In our proposed pre-ranking framework ASMOL, we inherit the features and model structures from the ranking models, as most pre-ranking works do. Meanwhile, we emphasize the importance of training sample mining which has yet to be noticed in previous pre-ranking works.

## 3 EVALUATION METRICS FOR PRE-RANKING

Considering the two goals mentioned in Section 1, we first analyze the existing metrics including AUC and $hitrate@k$ in Section 3.2 and Section 3.3, respectively. Furthermore, we introduce a novel offline metric called $ASH@k$ in Section 3.3. In Section 3.4, we analyze the current ability of each stage in Taobao Search using our new metric to demonstrate how to use the new metric to analyze a multi-stage e-commerce searching system.

### 3.1 Problem Formulation

In this section, we first formulate the pre-ranking problem and mathematical notations in the rest of this paper. Let $\mathcal{U} = \{u_1, \cdots, u_{|\mathcal{U}|}\}$ denotes the set of users together with their features. User features mainly consist of users' behavior information such as their clicked, collected, and purchased items and items added to the shopping cart. $Q = \{q_1, \cdots, q_{|Q|}\}$ denotes the set of search queries together with their corresponding segmentation. And $\mathcal{P} = \{p_1, \cdots, p_{|\mathcal{P}|}\}$ denotes the set of products (items) together with their features. Item features mainly consist of item ID, item statistical information, items' seller, etc. $|\mathcal{U}|, |Q|, |\mathcal{P}|$ are the numbers of distinct users, queries and items, respectively.

We combine each item $p_t$ in the matching output set with the user $u$ and the query $q$ as a triple $(u, q, p_t)$ when a user $u$ submits a query $q$. The pre-ranking models output the scores in each triple and usually select the top $k$ items from the matching output set according to the scores. Formally, given a triple $(u, q, p_t)$, the pre-ranking model predicts the score $z$ as follows:

$$z = \mathcal{F}(\phi(u, q), \psi(p)) \tag{1}$$

where $\mathcal{F}(\cdot)$ is the score function, $\phi(\cdot)$ and $\psi(\cdot)$ are the user and item embedding functions, respectively. In this paper, we follow the vector-product-based model framework [8] and adopt the cosine similarity operation as $\mathcal{F}(\cdot)$.

## 3.2 Consistency with the ranking stage

Considering the offline metrics of the ranking system in the industry, AUC is the most popular one. Taking Taobao Search as an example, AUC is computed over the exposures. As the goal of Taobao Search is to improve transaction efficiency, the ranking stage mainly considers the purchases as positives on exposures and cares about the likelihood that a purchased item is ranked higher than the others. As the maximum number in Taobao Search is set to 10 for each request, we use Purchase AUC at 10 ($PAUC@10$) to measure the model's capability for the ranking. As a result, $PAUC@10$ is conventionally used in pre-ranking, the same as that in ranking, and can measure the consistency with the online ranking system [11, 15, 24].

## 3.3 Quality of the output set

Recent pre-ranking works rarely use any metric to evaluate the quality of the whole output set. One evaluation metric for evaluating the output set is $hitrate@k$ (or $recall@k$), which is widely used in the matching stage. $hitrate@k$ denotes whether the model ranks the target (clicked and purchased) items within the top $k$ of the candidate set. Formally, for a $(u,q)$ pair, $hitrate@k$ is defined as follows:

$$hitrate@k = \frac{\sum_{i=1}^{k} \mathbb{1}(p_i \in T)}{|T|} \quad (2)$$

where $\{p_1, \cdots, p_k\}$ denotes the top $k$ items return by the pre-ranking model, $T$ denotes the target-item set which contains $|T|$ items, and $\mathbb{1}(p_i \in T)$ is 1 when $p_i$ is in the target set $T$, otherwise it is 0.

When using the metric in the matching stage, it is used to measure the quality of one matching model output set (*Matching1* in Figure 1, for example, not the entire output set of all the matching models online. In contrast, when we evaluate pre-ranking experiments via $hitrate@k$, the offline metrics' conclusion contradicts the online business metrics' conclusion. After further analysis, we found that selecting $k$ in the hitrate is a non-trivial problem. In order to accurately evaluate the quality of the output item set of the pre-ranking stage, $k$ is supposed to be equal to the size of the pre-ranking output set $|\mathcal{R}|$. However, since only items in the pre-ranking output set can be exposed and purchased during the online serving, all positives (target items) are in the pre-ranking's online output set. That causes the online pre-ranking model's $hitrate@k \equiv 1$ when $k = |\mathcal{R}|$. As a result, offline $hitrate@k$ can **only measure the difference between the offline models' output sets and the online output set rather than their quality.** The conventional pre-ranking methods [11, 24] use $k \ll |\mathcal{R}|$ to avoid the above problem. The disadvantage of $k \ll |\mathcal{R}|$ is obvious, as **it cannot indicate the quality of the entire pre-ranking output set.**

In this work, we propose a new effective evaluation metric called *All-Scenario Hitrate@k* (*ASH@k*). To create a metric truly indicating the quality of the pre-ranking output set, we introduce more positives (e.g., purchasing samples) from other scenarios in Taobao, such as recommendations, shopping carts, and ads. Since some positives from other scenarios do not exist in the pre-ranking online outputs, they can reveal users' preferences without in-scenario bias. In this case, $hitrate@k$ does not equal to 1 even if $k = |\mathcal{R}|$. As we care more about the transaction in Taobao Search, we only use purchasing positives from non-search scenarios. To distinguish between the hitrate with different positives, we call the $hitrate@k$ with purchasing samples only in searching as *In-Scenario Purchase Hitrate@k* (*ISPH@k*) and $hitrate@k$ with purchasing samples in all scenarios as *All-Scenario Purchase Hitrate@k* (*ASPH@k*).

Next, we detail how we introduce positives from other scenarios. One positive sample in evaluation is a triple with a user, a query, and an item: $(u_i, q_j, p_t)$. However, there is no corresponding query in most non-search scenarios like the recommendation scenario. To build evaluation samples for searching, we need to attach a non-search purchasing $(u_i, p_t)$ with a query requested by the corresponding user $(u_i, q_j)$. Suppose $A_u^i$ denotes the target-item set user $u_i$ purchases in all Taobao scenarios, $Q_u$ denotes all queries user search in Taobao search. An intuitive way is to build a Cartesian Product between queries and purchased items by the same user and use all triples $(u_i, q_j, p_t)$ as positive samples, where $q_j \in Q_u$ and $p_t \in A_u^i$. However, it will introduce a lot of irrelevant query-item pairs. For instance, one may search for "iPhone" in Taobao Search and buy some fruits in another recommendation scenario. The sample with "iPhone" as a query and "apple (fruit)" as an item is unsuitable for a positive sample in Taobao Search. In order to filter irrelevant samples, we only keep relevant samples whose relevance score of $(q_j, p_t)$ is above the borderline[1]. We call the $q_k$ a "related query" for the all-scenario pair $(u_i, p_t)$; meanwhile, the $p_t$ is an all-scenario "related item" that can be attached to the in-scenario pair $(u_i, q_j)$. Furthermore, we also removed duplicate samples. Each $(u,p)$ pair in the triple is unique, so even if $u_i$ purchased a $p_t$ more than once, we only evaluate the triple with $(u_i, p_t)$ only once. Meanwhile, if we can find more than one related query before the user's purchase behavior and construct different triples like $(u_i, q_1, p_t)$, $(u_i, q_2, p_t)$ and $(u_i, q_j, p_t)$, we only keep the triple with latest $q$ for evaluation. Formally, similar to Eq.2, for each $(u_i, q_k)$ pair, $ASPH@k$ is defined as:

$$ASPH@k = \frac{\sum_{i=1}^{k} \mathbb{1}(p_i \in A_u^i)}{|A_u^i|} \quad (3)$$

where $A_u^i$ denotes the target-item set containing $|A_u^i|$ items that $u_i$ purchases in any scenario and can be attached to the query.

## 3.4 All-scenario purchase hitrate of each stage in Taobao Search

We show the offline metrics of the pre-generation pre-ranking model, the proposed pre-ranking model, and the ranking model on the Taobao Search [2] in Figure 2. To fairly compare the model ability in the pre-ranking stage, all of these models are evaluated on the pre-ranking candidates. For the pre-generation pre-ranking model that uses the same sample as the ranking model, its model ability is weaker than the ranking model from $10^5$ to $10^1$ consistently. By contrast, the proposed pre-ranking model significantly outperforms that of ranking on both *ASPH@k* and *ISPH@k* when $k$ gets larger. **This phenomenon implies that the proposed pre-ranking model ability can outperform the ranking when outputting thousands of items.**

---

[1]The borderline is defined by the relevance team, which is supposed to guarantee the user's experience by not showing the irrelevant items.

[2]the dataset is sampled from the total real-world dataset, which has the same pattern and characteristic as the total Taobao Search online dataset. However, due to the data security, we only display the absolute values in the subset since it has the same regularity as the complete dataset.

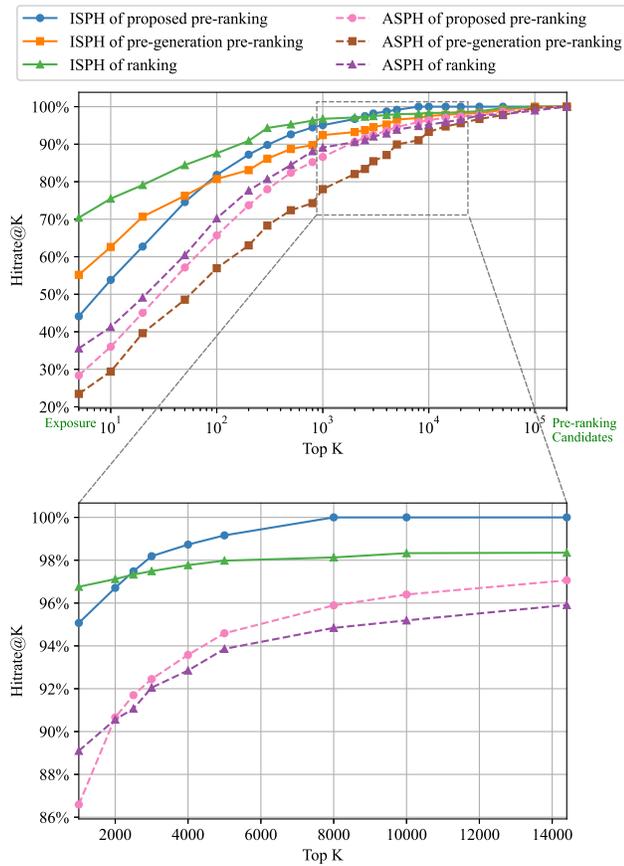

**Figure 2: Hitrate@$k$ in Taobao Search. The figure above is zoomed as the figure below.**

Meanwhile, there is a significant difference between the results of *ASPH@k* and the results of *ISPH@k* in Figure 2. From the view of the *ISPH@k* metric, the ranking model outperforms the pre-ranking model when $k$ is less than 3000, while from the view of *ASPH@k* metric, it only outperforms the pre-ranking model when $k$ is less than 2000 approximately. As mentioned in Section 3.3, we argue that the *ISPH@k* scores indicate the differences between the offline and online sets and cannot necessarily indicate the offline set's quality. Because the ranking model's scores determine the final exposed items, the ranking model will have a significant advantage when using *ISPH@k* as an evaluation metric.

To further validate the effectiveness of *ASPH@k*, we perform an online A/B test with the pre-ranking outputting 2500 and 3000 items, respectively. If the *ISPH@k* evaluation is effective, then the online business metric of the pre-ranking outputting of 3000 items should be higher. If the *ASPH@k* is effective, then the conclusion is the opposite. The online results show that the pre-ranking outputting

2500 has a 0.3% online 30-day A/B transaction GMV improvement compared to 3000. **This experiment verifies that the *ASPH@k* is a more reliable metric than the *ISPH@k* in offline evaluation.** Moreover, the experiment also shows that the pre-ranking can possess the ability that the ranking does not possess, as the size of the pre-ranking output set is not the larger the better. Consequently, a pre-ranking should develop its advantage on higher quality outputs rather than imitate ranking blindly.

## 4 ON OPTIMIZATION OF THE PRE-RANKING

Although imitating the ranking makes a pre-ranking more consistent with the ranking, it can contribute little to the quality of a whole pre-ranking's output set. This section discusses our optimization techniques to reach the pre-ranking goals, including training data construction, all-scenario labels, loss functions, and distillation. We first introduce the overall framework of the proposed All-Scenario based Multi-Objective Learning (ASMOL) pre-ranking in Section 4.1, then analyze each part with ablation studies in Section 4.2 - 4.5. The experimental results not only prove the effectiveness of our proposed optimization method but also validate the consistency of the proposed *ASPH* with online business metrics. Besides, we also discuss the combination strategies for multi-objectives and the related experiments in Section 4.6.

### 4.1 Overall framework for multi-objective learning

In order to improve both the quality of the pre-ranking output set and the consistency with the ranking, we design a novel All-Scenario-based Multi-Objective Learning framework. On the one hand, we extend the training sample from exposures to an entire-space training sample with the all-scenario label, which mainly aims to improve the quality of the pre-ranking output set. On the other hand, we design a distillation loss to improve consistency with the ranking stage. Figure 4 and Figure 3 show the training framework of the existing pre-ranking models and our pre-ranking model, respectively.

Figure 4 shows the common framework of the traditional pre-ranking. The training sample in a traditional pre-ranking model [8] consists of features of a user, a query, and an item in exposures. The labels for Click Through Rate (CTR) task and Conversion Rate (CVR) task are in-scenario click and purchase. This pre-ranking system consists of two different models predicting the CTR and CVR, and the final efficiency score[3] is CTR*CVR. As the previous generation pre-ranking model in Taobao Search followed this framework which outputs the CTR score and the CVR score separately, we call the framework in Figure 4 as *baseline* for convenience.

In contrast, as is shown in Figure 3, our ASMOL is trained on query-level training samples with multiple objectives and multiple positive items. Specifically, there are features of the user, query, and three kinds of items in each training sample. All items share the same deep neural network with the same trainable parameters. The input features and model structures for processing each user, query, and item are illustrated in Appendix A, which are not significantly different from existing pre-ranking models. Each sample contains all items exposed in a query (request) of Taobao Search and candidates of ranking and pre-ranking sampled from the online log system. The relationship of three kinds of training samples is illustrated in Figure 5, and the definition is as follows:

- **Exposures (Ex, also called impressions)**: $N$ items exposed in a query (request), including the items clicked or purchased by the user. These presented items were sorted at the top of all candidates by the cascade ranking system and are predicted to have the largest probabilities to satisfy the user.
- **Ranking Candidates (RC)**: Items output by the online pre-ranking and served as the ranking candidates, which were not exposed to the user. We sample $M$ items from thousands of ranking candidates for each query through an online logging system.
- **Pre-Ranking Candidates (PRC)**: Items not output by the pre-ranking system. We sample $L$ items from hundreds of thousands of items for each query through an online logging system.

---

[3]As they are different score in Taobao Search including efficiency score, price score, relevance score,etc, we only considering efficient in this work because others' settings are the same.

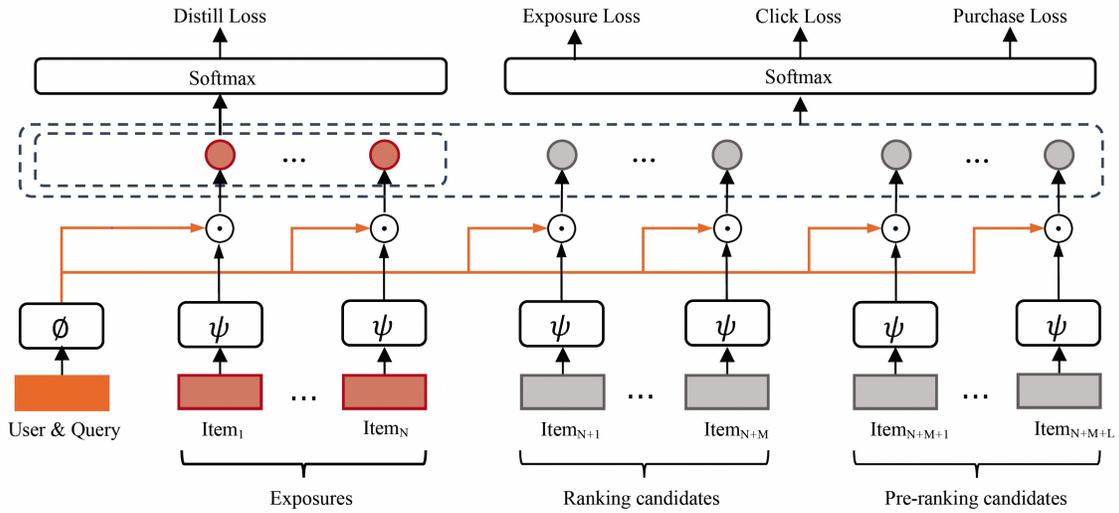

**Figure 3: The All-Scenario-based Multi-Objective Learning framework (ASMOL) in Taobao Search**

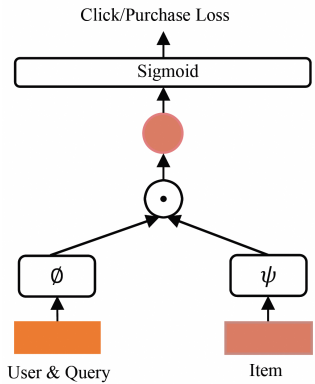

**Figure 4: The common training framework for pre-ranking, which is called *baseline* in this work.**

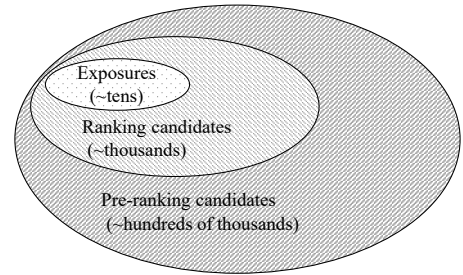

**Figure 5: Training samples of our model. Note that exposures are excluded from the ranking candidates. Both exposures and ranking candidates are excluded from the pre-ranking candidates. There is no intersection between the three sample sets. The training samples are sampled from these three sets.**

Both Ranking Candidates and Pre-ranking Candidates are negatives in all the objectives. And both types of negatives aim to resolve the SSB problem in the current pre-ranking system. Furthermore, the Pre-ranking Candidates are trained as easy examples, and Ranking Candidates are relatively hard examples, inspired by the strategy used in matching system [7]. We demonstrate the necessity of different samples in Section 4.2.

Moreover, we also adopt three kinds of binary labels corresponding to three optimization objectives for all items in each sample:

- **All-Scenario Purchase Label (ASPL)**: This label denotes whether the user in any scenario purchases the item. If the user purchased the item in the current query (of Taobao Search) or the user purchased the item in other scenarios and can be identified as a related item to the current query, the purchase label is 1. Otherwise, the purchase label is 0.

- **All-Scenario Click Label (ASCL)**: Similar to ASPL, the label denotes whether the user in any scenario clicks the item. As the user has to click an item before he purchases it, but the click and purchase behavior can happen in different scenarios that may result in conflict labels, we set the item's click label as 1 whenever its ASPL is 1.
- **Adaptive Exposure Label (AEL)**: This label denotes whether the item is exposed in this request of Taobao Search, which is determined by the cascade ranking system. The labels are 1 for exposures and 0 for all other items. Besides, we set the item's exposure label as 1 whenever its ASCL is 1.

To reach the two goals of pre-ranking, we create a loss function combining a ranking loss and a distillation loss:

$$L = L_{\text{rank}} + L_{\text{distill}} \qquad (4)$$

The ranking loss $L_{\text{rank}}$ utilizes the three different labels simultaneously. We create a multi-objective loss function with three tasks: Purchase, Click, and Exposure. We adopt a new list-wise loss for each task to make the logit of the positive sample larger than the negative sample. Our multi-object learning framework aims to learn the following order of importance: purchased items > clicked but non-purchased items > non-click exposures > ranking candidates (RC) and pre-ranking candidates (PRC) by maximizing the scores of different types of positive items. We demonstrate the necessity of multi-objective optimization in Section 4.3. The list-wise loss is discussed in Section 4.4.

Moreover, we add an auxiliary distillation loss $L_{\text{distill}}$ to learn from the ranking model with more features and trainable parameters. Surprisingly, we find that simply distilling all training samples is not the best solution. The ranking model is not always a good teacher, especially in samples that have not been exposed. We analyze this phenomenon in Section 4.5.

### 4.2 Entire-space training samples

**Table 1: Experiment results for different sample strategies**

| Sample strategy | *ASPH*@3000 | *PAUC*@10 | GMV gained |
| --- | --- | --- | --- |
| *baseline* | 85.5% | 90.1% | 0.0% |
| ASMOL | 92.5% | 87.1% | 1.2% |
| ASMOL w/o PRC | 91.1% | 88.9% | 1.0% |
| ASMOL w/o RC | 90.4% | 87.5% | 0.8% |
| ASMOL w/o RC&PRC | 87.3% | 89.4% | 0.3% |

As illustrated in Table 1, when the training samples include the exposures, the ranking candidates, and pre-ranking candidates, both the *ASPH*@3000 and online GMV are improved. If the pre-ranking model is only fed with exposures, we find that the quality of its output set is terrible by analyzing the cases sampled from the online log system. Since the exposure-sample-only pre-ranking model did not see non-exposed samples during training, it gets confused when given samples in the pre-ranking candidates in evaluation. In this way, the scores for most candidates are not convincing, resulting in many low-quality items in the output set.

We further explore the impact of the proportion of different samples. As some researchers point out, there is usually an optimal ratio of different types of negative samples[7] and an optimal ratio of negatives and positives[29]. In our multi-objective framework, taking *click* objective for example, items whose ASCL equal 1 are positives, and others are negatives. The negative samples from exposures, ranking candidates to pre-ranking candidates, become increasingly easy. Similar to the click objective, the purchase objective also contains these three types of negatives. However, for the exposure objective, all the exposure samples are positive, and ranking candidates and pre-ranking candidates are negative. As a result, we find that even if we removed a proportion of the non-click exposures, it harmed the exposure objective and made the *ASPH*@3000 drop significantly. Meanwhile, the number of RC and the number of PRC sampled from the online log system are not the larger, the better. The details are shown in Figure 6 and the y-axis indicates the offline metrics' gaps adding the different number of candidates based on the experiment "ASMOL w/o RC&PRC" shown in Table 1. In order to maximize the *ASPH*@3000, we set the number of RC and the number of PRC as 10 and 40, respectively. Besides, we can see a clear "Seesaw Effect" when RC and PRC is smaller than 10 and 40, respectively, probably because the model has reached the Pareto frontier.

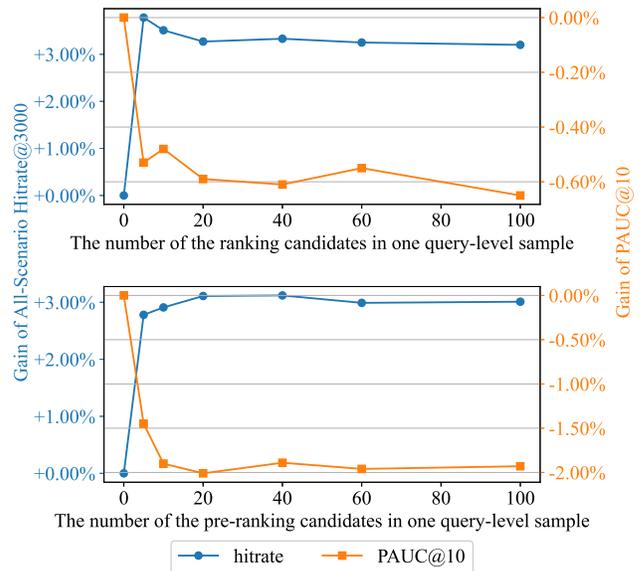

**Figure 6: Experiment results on different sample sizes. The baseline is the ASMOL without both ranking cadidates (RC) and pre-ranking cadidates (PRC).**

Moreover, as the pre-ranking uses RC and PRC while the ranking doesn't, it makes the pre-ranking different from the ranking. It naturally makes the pre-ranking outperform the ranking when outputting thousands of items, as shown in Section 3.4 Figure 2. The online A/B test results also show that when the conclusion between *ASPH*@3000 and *PAUC*@10 conflicts, *ASPH*@3000 is more reliable, which is more consistent with the pre-ranking's primary goal and online business metrics.

## 4.3 All-scenario labels in multi-objective learning

**Table 2: Experiment results for different label strategies**

| label strategy | $ASPH@3000$ | $PAUC@10$ | GMV gained |
|---|---|---|---|
| ASMOL | 92.5% | 87.1% | 1.2% |
| ASMOL w/o AEL | 90.2% | 87.9% | 0.8% |
| ASMOL w/o ASCL | 91.5% | 87.4% | 1.0% |
| ASMOL w/o ASPL | 91.2% | 85.4% | 0.8% |
| ASMOL w/ ASPL->ISPL | 90.8% | 87.3% | 0.9% |
| ASMOL w/ ASCL->ISCL | 91.6% | 87.2% | 1.0% |
| ASMOL w/ ASL->ISL | 89.7% | 87.4% | 0.7% |

We conduct an ablation study to investigate the effectiveness of all-scenario labels used in our multi-objective learning framework by removing each label's corresponding loss separately. The experimental results are reported in the first four-line results of Table 2. When we remove the adaptive exposure label (AEL), all-scenario click label (ASCL) and all-scenario purchase label (ASPL) by removing their corresponding loss respectively, the results of evaluation metrics show that the model performances drop significantly. The exposure label helps the pre-ranking model learn the ranking patterns of the downstream cascade ranking system, and our results validate its effectiveness. Meanwhile, the ASCL and ASPL have less in-scenario bias than AEL and can give the pre-ranking model more precise information about users' preference. The experimental results prove that the combination of our three losses is reasonable and effective in improving the online and offline metrics of pre-ranking models.

We also conduct several experiments on the effect of All-Scenario Label (ASL) compared with In-Scenario Label (ISL), which is shown in the last three-line results in Table 2. The symbol "->" denotes changing one label to another. For example, "ASL->ISL" denotes that both ASPL and ASCL are changed to ISPL and ISCL, respectively. By comparing the results between ASMOL with the last three lines in Table 2, we can naturally conclude that utilizing ASL is more consistent with $ASH@k$ and online business metrics. Meanwhile, the AUC may drop due to the "Seesaw Effect".

## 4.4 List-wise loss for multi-positive label

We design a multi-objective loss to simultaneously combine exposure, click, and purchase in one model. These three optimization goals are trained jointly:

$$L_{rank} = \alpha_{ex} L_{exposure} + \alpha_{cl} L_{click} + \alpha_{pur} L_{purchase} \quad (5)$$

For each task, we use the list-wise ranking loss [17, 29]. For example, for the purchasing task, the ranking loss can be formulated as follows:

$$L_{purchase} = \sum_{i \in \mathcal{D}} -\log \frac{\exp z_i}{\sum_{j \in \mathcal{S}} \exp z_j} \quad (6)$$

where $z$ is the logit, $\mathcal{S}$ is the full training sample set, and $\mathcal{D}$ is the positive sample set of the purchase task, including exposures, ranking candidates, and pre-ranking candidates as discussed in Section 4.1. Eq.6 works well for the purchasing task because there is usually at most one purchasing in one query. However, the multi-positive label is common in clicking and exposure tasks. Eq.6 is no longer proper for tasks with multiple positive samples [20]. For a multi-positive task, the vanilla Softmax in Eq.7 leads the optimization to fall into the comparison between positive samples, rather than the comparison between positive and negative samples. Inspired by CircleLoss [22], we slightly modify the Eq.6 for tasks with multiple positive labels:

$$L_{exposure} = \sum_{i \in \mathcal{E}} -\log \frac{\exp z_i}{\sum_{j \in \{\mathcal{S} \setminus \mathcal{E}, i\}} \exp z_j} \quad (7)$$

where $\mathcal{E}$ denotes the training sample set of the exposure task. Eq.7 degenerates into Eq.6 when there is only one positive sample. We will detail Eq.7 in Appendix B. In order to verify the effectiveness of the modification in the Eq.7, we conduct the experiments to compare these two loss functions and show results in table 3. The experiment unsurprisingly shows that the multi-label Softmax outperforms the vanilla Softmax in all metrics. We tune weights of each task $\alpha_{ex}$, $\alpha_{cl}$ and $\alpha_{pur}$ following Zheng et al. [29].

**Table 3: Experiment results on different loss functions.**

| Loss function | ASPH@3000 | PAUC@10 | GMV Gained |
|---|---|---|---|
| ASMOL w/ -vanilla Softmax | 92.0% | 85.9% | 1.0% |
| ASMOL (Eq.7) | 92.5% | 87.1% | 1.2% |

## 4.5 Distillation from the ranking stage

It is natural to train a compact pre-ranking model by distilling it from the ready-made large-ranking model with more features and parameters. In the Taobao Search ranking stage, there are two separate models that correspond to CTR and CVR predictions. In the pre-ranking stage, we use the calibrated[4] CTR and CVR predictions as teachers to distill our pre-ranking model. In order to fully utilize both CTR and CVR model, the pre-ranking distillation loss is the combination of the CTR distill, and the click through rate & conversion rate (CTCVR) distill:

$$L_{distill} = \alpha_{cl} L_{CTR} + \alpha_{pur} L_{CTCVR} \quad (8)$$

The CTR distillation task can be formulated as follows:

$$L_{CTR} = \sum_{i \in \mathcal{D}} -p_{CTR} \log \frac{\exp z_i}{\sum_{j \in \mathcal{D}} \exp z_j} \quad (9)$$

where $z$ is the logit of the pre-ranking model, $\mathcal{D}$ is the sample set to be distilled, and the teacher $p_{CTR}$ is the prediction of the ranking CTR model. Similarly, the teacher of the CTCVR distillation is $p_{CTR} * p_{CVR}$. Both $p_{CTR} = p(click|expose)$ and $p_{CTCVR} = p(click\&purchase|expose)$ are the conditional probability given the same condition $p(expose) = 1$. Since the condition of $p_{CVR} = p(purchase|click)$ is different from $p_{CTR}$, we use $p_{CTCVR}$ as the soft label to be consistent with $p_{CTR}$. The loss weight of the CTR distillation and the CTCVR distillation follow the weights of the click task and the purchase task, respectively.

---

[4]Both CTR and CVR predicted by the ranking system are well calibrated on impressions so that they can indicate the probability of click and conversion, respectively.

Furthermore, we observe that defining the distillation sample set $\mathcal{D}$ is a non-trivial problem. The straightforward way is distilling all training samples in the pre-ranking, i.e., $\mathcal{D} = \mathcal{S}$. Since the ranking is trained with only exposures, the ranking is not expert on non-exposed samples. Hence, we need to select which samples to distill from the ranking model carefully. We set up different experiments in the pre-ranking to validate the influence of different distillation strategies. As is shown in 4, when we extend the distillation sample set from $Ex$ to $Ex+RC$, the $PAUC@10$ is improved as the pre-ranking becomes more consistent with the ranking. However, the offline metric $ASPH@3000$ and online GMV drops because the ranking is not a good teacher of non-exposed samples. In contrast, by comparing the results between $Ex(ASMOL)$ with $Ex(no\ distillation)$, we know that the ranking is a good teacher on Ex. Only learning the ranking scores on Ex with proper loss weights will help the pre-ranking reach the Pareto frontier without causing the "Seesaw Effect". The experiment results also verified our conclusion shown in Section 3.4: **the pre-ranking can not blindly follow the ranking.**

**Table 4: Experiment results on distillation with different samples.**

| Samples to distill | ASPH@3000 | PAUC@10 | GMV Gained |
|---|---|---|---|
| Ex (ASMOL) | 92.5% | 87.1% | 1.2% |
| Ex+RC | 92.0% | 87.4% | 1.1% |
| no distillation | 92.5% | 85.1% | 0.9% |

### 4.6 Combination strategies for multi objectives

The previous version of pre-ranking systems in Taobao Search contains two models that output CTR and CVR separately, and the final efficient score is $p_{CTR} * p_{CVR}$. We call this combination strategy between different tasks a multi-model strategy, and call our ASMOL using only one model to learn different tasks a one-model strategy. We perform several experiments to compare the different combination strategies for multi-objectives fairly. We perform experiments over different training sample sets. The three experiments shown in Table 5 are training only on exposures. The *baseline* denotes the previous version pre-ranking model, which belongs to the one-model strategy and outputs the CTR score and CVR score separately via two different models. We set up experiments of "*baseline* w/ ASL&distill" by adding the distill loss and changing the label of its CTR and CVR task from origin In-Scenario Label (ISL) to All-Scenario Label (ASL), respectively, in order to make it can be fairly compared with "ASMOL w/o RC&PRC". As a result, we can see that the offline and online metrics of "ASMOL w/o RC&PRC" is higher than that of "*baseline* w/ ASL&distill", which indicates that the one-model strategy outperforms the multi-model strategy. When extending the training set from EX to the entire training space consisting of Ex, RC, and PRC, we add a new model to output $p_{ER} = p(\text{expose}|\text{pre} - \text{ranking candidates})$ denoting the exposing rate from pre-ranking candidates. The ER model is the same as the exposure task in ASMOL, and the final efficient score is $p_{CTR} * p_{CVR} * p_{ER}$. By comparing the results of "*baseline* w/ ASL&distill&ER" with "ASMOL", we can see that the $ASPH@3000$

and online business metric of our one-model strategy is still higher than that of the three-model strategy. In conclusion, our ASMOL is more effective than the multi-model strategy though the multi-model strategy uses double or triple parameters.

Eventually, comparing the results between "*baseline*" and "AS-MOL" shows that our new framework has significantly improved the offline metrics $ASPH@3000$ and improved 1.2% online GMV.

**Table 5: Experiment results for different combination strategies training only on exposures(Ex).**

| Strategy | ASPH@3000 | PAUC@10 | GMV gained |
|---|---|---|---|
| *baseline* | 85.5% | 90.1% | 0.0% |
| *baseline* w/ ASL&distill | 86.5% | 89.9% | 0.2% |
| ASMOL w/o RC&PRC | 87.3% | 89.4% | 0.3% |

**Table 6: Experiment results for different combination strategies training on samples consisting of Ex,RC and PRC.**

| Strategy | ASPH@3000 | PAUC@10 | GMV gained |
|---|---|---|---|
| *baseline* w/ -ASL&distill&ER | 89.6% | 88.6% | 0.7% |
| ASMOL | 92.5% | 87.1% | 1.2% |

## 5 CONCLUSION AND FUTURE WORK

In conclusion, the pre-ranking play a unique role in outputting an optimal high-quality item set in the e-commerce searching system. To improve the online business metrics, we should focus more on the quality of the unordered output set when optimizing the pre-ranking. On the one hand, researchers need to use the $ASH@k$ we proposed to measure the quality of the unordered set, which can save much more time than online experiments. On the other hand, the pre-ranking model should not blindly follow the ranking model because the ranking is not aways a good teach, and it can cause the "Seesaw Effect" when maximizing the consistency with the ranking. Researchers need to analyze both the ranking and the pre-ranking through $ASH@k$ in fair, and decide from which scores the pre-ranking can learn. The ASMOL framework proposed in our work achieves great success. Because it not only improves 1.2% online GMV on Taobao Search but also proves that the light pre-ranking model can outperform the large ranking model when outputting thousands of items, as shown in Figure 2 of Section 3.4. In the future, from the view of $ASH@k$, Taobao Search needs to change the ranking system's learning framework to perform better on its candidate set and get more consistent with the pre-ranking.

# Appendices

## A  INPUT FEATURES AND MODEL STRUCTURE IN PRE-RANKING SYSTEM

Figure 7 shows the architecture of the pre-ranking model in detail. The model generates the user-query embedding and the item embedding independently and finally performs the inner product between them as the prediction score.

### A.1  User-Query Tower

The user-query tower consists of three units: query semantic unit, user behavior attention unit, and embedding prediction unit.

*A.1.1  Query Semantic Unit.* With the embedding matrix of the query term sequence $e^q = \{e^{w_1}, \cdots, e^{w_{|q|}}\} \in \mathbb{R}^{|q| \times d}$ as the input, the query semantic unit generates three kinds of query representations: mean-pooling representation $Q_m \in \mathbb{R}^{1 \times d}$, self-attention representation $Q_s \in \mathbb{R}^{1 \times d}$ and personalized representation $Q_p \in \mathbb{R}^{1 \times d}$. The overall query representation $Q_o \in \mathbb{R}^{1 \times 3d}$ is obtained as follows:

$$Q_o = concat(Q_m, Q_s, Q_p) \tag{10}$$

$$Q_m = mean\_pooling(e^q) \tag{11}$$

$$Q_s = max\_pooling(self\_atten(e^q))$$

$$= max\_pooling(\text{Softmax}(\frac{e^q \cdot (e^q)^T}{\sqrt{d}}) \cdot e^q) \tag{12}$$

$$Q_p = \text{Softmax}(\frac{(e^u W_1 + b_1) \cdot (e^q)^T}{\sqrt{d}}) \cdot e^q \tag{13}$$

where *mean_pooling*, *max_pooling* and *concat* are the average, maximum and concatenation operation, and $e^u$ is the user representation obtained by concatenating all embeddings of the user profile features, $W_1 \in \mathbb{R}^{d \times d'}$ and $b_1 \in \mathbb{R}^{1 \times d'}$ are the parameters of a fully connected layer for $e^u$. With this query semantic unit, we fuse generic and personalized representations of the query, which helps capture rich semantic information in the query.

*A.1.2  User Behavior Attention Unit.* Let $\mathcal{B} = \{p_1^u, \cdots, p_{|\mathcal{B}|}^u\}$ denote the historical behaviors of the user $u$, including $u$'s clicked, collected, and purchased items as well as items added to the shopping cart. Further, we divide the user's behaviors into three non-overlapping collections according to the time interval from the current time: real-time behaviors $\mathcal{B}_r$ in the past one day, short-term behaviors $\mathcal{B}_s$ from the past second day to the past tenth day, and long-term behaviors $\mathcal{B}_l$ from the past eleventh day to the past one month, i.e., $\mathcal{B} = \mathcal{B}_r \bigcup \mathcal{B}_s \bigcup \mathcal{B}_l$. Users' historical behaviors reflect their rich interests. However, too many noisy behaviors harm the model's performance. To remove the impact of these noisy behaviors, we first train a deep learning model to predict the relevant categories of queries and then filter out the historical behaviors that do not belong to the predicted relevant categories of the query. This operation is called *Category Filtering* [27]. Next, we adopt the query attention mechanism [10] to mine useful information from the reserved historical behaviors effectively. We use the embeddings of user profile features and query features as the *Query* of the attention unit and the user's historical behavior sequence as the *Key* and *Value*. We perform the same query attention on three collections of user behaviors $\mathcal{B}_r$, $\mathcal{B}_s$, and $\mathcal{B}_l$. Taking $\mathcal{B}_r$ for example, each item $p \in \mathbb{R}^{1 \times d_i}$ in $\mathcal{B}_r$ is represented by concatenating its ID and side information embeddings together along the last axis. Next, the embedding matrix of the real-time behavior sequence denoted as $\mathbb{B}_r = (p_1, \cdots, p_{|\mathcal{B}_r|}) \in \mathbb{R}^{|\mathcal{B}_r| \times d_i}$, is aggregated by the query attention. The output $H_r$ of the query attention is defined as follows:

$$H_r = \text{Softmax}(\frac{Q \cdot (\mathbb{B}_r)^T}{\sqrt{d_i}}) \cdot \mathbb{B}_r \tag{14}$$

$$Q = concat(Q_o, e^q, e^u) W_r + b_r \tag{15}$$

where $Q_o$, $e^q$, $e^u$ are the query semantic representation generated from the query semantic unit, the concatenated embeddings of query side information (such as query frequency and relevant categories), and the concatenated embeddings of the user profile features, respectively. $W_r \in \mathbb{R}^{d_{uq} \times d_i}$ and $b_r \in \mathbb{R}^{1 \times d_i}$ are the parameters of a fully connected layer for the real-time behavior sequence.

Next, we concatenate the representations of real-time, short-term, and long-term behaviors, i.e., $H_r$, $H_s$, and $H_l$, as the final representation of user historical behaviors $H_B$:

$$H_B = concat(H_r, H_s, H_l) \tag{16}$$

*A.1.3  Embedding Prediction Unit.* After processing all query and user features, we concatenate all representations of user and query together and feed the new representation $e^{uq}$ into a Multilayer Perceptron ($MLP_{uq}$) with four layers. Each layer of the first three layers consists of a fully connected layer ($FC$), a layer normalization layer ($LN$), and a Leaky ReLU layer ($LReLU$). The last layer of MLP is a fully connected layer ($FC$). The user-query representation $H_{uq}$ is obtained by normalizing the output vector of MLP to a unit vector:

$$H_{uq} = l2\_norm(MLP_{uq}(e^{uq})) \tag{17}$$

$$e^{uq} = concat(e^u, e^q, Q_o, H_B) \tag{18}$$

$$MLP_{uq} = \{(FC, LN, LReLU) \times 3, FC\} \tag{19}$$

### A.2  Item Tower

In the item tower, we first get the item embedding $e^p$ by concatenating all embeddings of item ID and item side information together. For the item title feature, we utilize the mean-pooling operation on embeddings of title terms to get its embedding. Then we feed the item embedding $e^p$ into another Multilayer Perceptron ($MLP_p$) with four layers. Thus, the final item representation $H_p$ is defined as:

$$H_p = l2\_norm(MLP_i(e^p)) \tag{20}$$

$$MLP_p = \{(FC, LN, LReLU) \times 3, FC\} \tag{21}$$

## B  LIST-WISE RANKING LOSS WITH MULTIPLE POSITIVE LABELS

Learning to rank (LTR) is an approach to learning samples' rank. The designation of the loss function is an active area in the metric learning[22]. The cross-entropy loss with Softmax is a common function in the cascade ranking system[1, 7, 19]:

$$L = \sum_{i \in \mathcal{S}} (-y_i \log(\text{Softmax}(z))) = \sum_{i \in \mathcal{S}} (-y_i \log(\frac{\exp z_i}{\sum_{j \in \mathcal{S}} \exp z_j})) \tag{22}$$

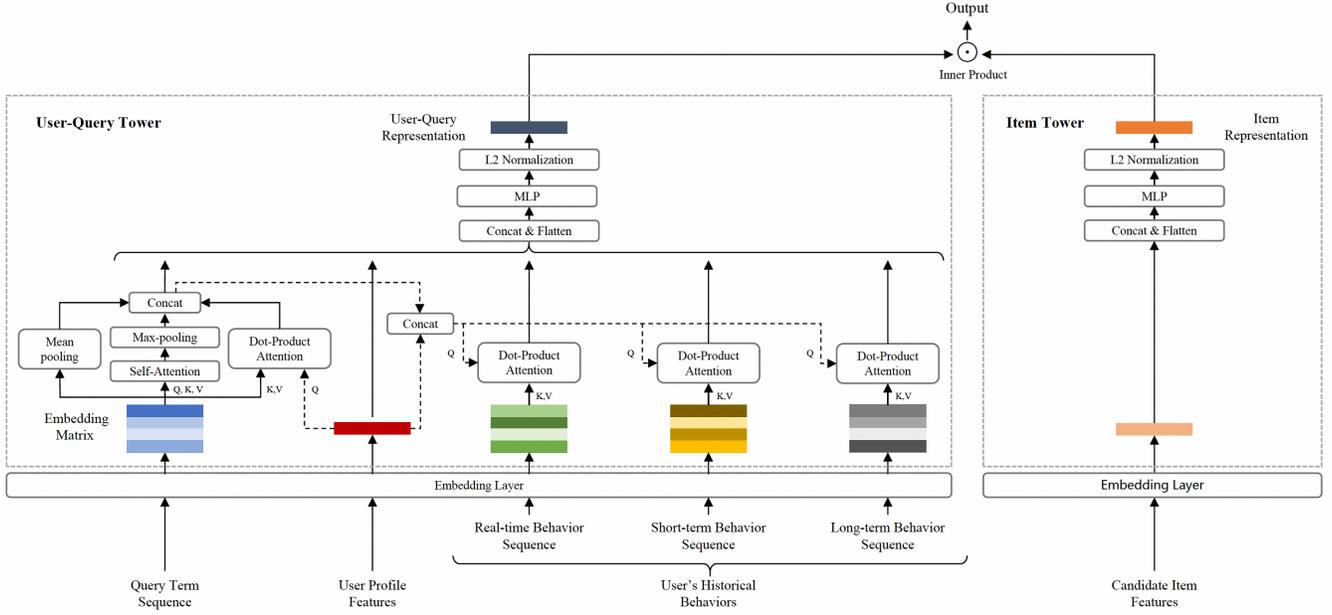

**Figure 7: The architecture of our pre-ranking model**

where $\mathcal{S}$ is a set of sample list, $\boldsymbol{y}$ is the label of each item, and $z$ is the logit of the item. The logit denotes the similarity between the user and the item in the ranking. The cross-entropy loss with Softmax in the Eq.22 aims to minimize the similarity score $z$ between positive samples $\mathcal{D}$ and negative samples $\mathcal{N}$. To understand how Eq.22 works, we rewrite it as follows:

$$
\begin{aligned}
L &= \sum_{i \in \mathcal{D}} \log \left( 1 + \exp \left( -z_i + \log \sum_{j \in \mathcal{S} \backslash i} e^{z_j} \right) \right) \\
&= \sum_{i \in \mathcal{D}} \text{SoftPlus} \left( \text{LogSumExp}_{j \in \mathcal{S} \backslash i} (z_j) - z_i \right) \\
&\approx \sum_{i \in \mathcal{D}} \left[ \max_{j \in \mathcal{S} \backslash i} (z_j) - z_i \right]_+ \\
&= \sum_{i \in \mathcal{D}} \left[ \max_{j \in \{ \mathcal{N} \cup (\mathcal{D} \backslash i) \}} (z_j) - z_i \right]_+
\end{aligned}
\tag{23}
$$

where

$$
\text{LogSumExp}(\boldsymbol{x}; \gamma) = \frac{1}{\gamma} \log \sum_i \exp(\gamma x_i) \approx \max(\boldsymbol{x})
\tag{24}
$$

and

$$
\text{Softplus}(\boldsymbol{x}) = \log(1 + e^{\boldsymbol{x}}) \approx \max(\boldsymbol{x}, 0) = [\boldsymbol{x}]_+
\tag{25}
$$

Eq.23 demonstrates the purpose of the cross entropy with softmax. For the single-positive sample task, the loss function minimizes the similarity *between the positive sample and the negative samples* by minimizing $L$ in Eq.23. However, the optimization goal changes when there are multiple positive samples. Since the scores $z$ of the positive samples are usually larger than the score of the negative samples, i.e., $\max_{j \in \mathcal{D}} z_j >= \max_{j \in \mathcal{N}} z_j$, we have:

$$
L \approx \sum_{i \in \mathcal{D}} \left[ \max_{j \in \{ \mathcal{D} \backslash i \}} (z_j) - z_i \right]_+
\tag{26}
$$

Eq.26 shows that the optimization goal turns out to be minimizing the similarity *in the positive samples*. The proposed Eq.7 solves the problem by removing the positives in the denominator of softmax. The modified loss function turns out to be:

$$
L \approx \sum_{i \in \mathcal{D}} \left[ \max_{j \in \mathcal{N}} (z_j) - z_i \right]_+
\tag{27}
$$

which avoids the unnecessary comparison between positive samples.